\documentclass[nonacm,authorversion,preprint,sigconf]{acmart}   

\graphicspath{{./}{../}} 

\AtBeginDocument{%
  }

\setcopyright{acmlicensed}
\copyrightyear{2024}
\acmYear{2024}
\acmDOI{XXXXXXX.XXXXXXX}

\acmConference[CHI]{CHI 2025}{April 26 -- May 01,
  2025}{Yokohama, JP}
\acmISBN{978-1-4503-XXXX-X/18/06}

\usepackage{multirow}
\usepackage{ccicons}
\usepackage[capitalise,nameinlink,noabbrev]{cleveref}

\begin{document}
\copyrightyear{2025}
\acmYear{2025}
\setcopyright{acmlicensed}\acmConference[CHI '25]{CHI Conference on Human Factors in Computing Systems}{April 26-May 1, 2025}{Yokohama, Japan}
\acmBooktitle{CHI Conference on Human Factors in Computing Systems (CHI '25), April 26-May 1, 2025, Yokohama, Japan}
\acmDOI{10.1145/3706598.3713949}
\acmISBN{979-8-4007-1394-1/25/04}
\title[Traversing Dual Realities]{Traversing Dual Realities: Investigating Techniques for Transitioning 3D Objects between Desktop and Augmented Reality Environments}


\author{Tobias Rau}
\orcid{0000-0002-3310-9163}
\affiliation{%
  \institution{University of Stuttgart}
  \city{Stuttgart}
  \country{Germany}}
\email{tobias.rau@visus.uni-stuttgart.de}

\author{Tobias Isenberg}
\orcid{0000-0001-7953-8644}
\affiliation{%
  \institution{Université Paris-Saclay, CNRS, Inria, LISN}
  \city{Orsay}
  \country{France}}
\email{tobias.isenberg@inria.fr}

\author{Andreas K{\"o}hn}
\orcid{0000-0002-0844-842X}
\affiliation{%
  \institution{University of Stuttgart}
  \city{Stuttgart}
  \country{Germany}}
\email{koehn@theochem.uni-stuttgart.de}

\author{Michael Sedlmair}
\orcid{0000-0001-7048-9292}
\affiliation{%
  \institution{University of Stuttgart}
  \city{Stuttgart}
  \country{Germany}}
\email{Michael.Sedlmair@visus.uni-stuttgart.de}

\author{Benjamin Lee}
\orcid{0000-0002-1171-4741}
\affiliation{%
  \institution{JPMorganChase}
  \city{New York}
  \country{USA}}
\affiliation{%
  \institution{University of Stuttgart}
  \city{Stuttgart}
  \country{Germany}}
\email{benjamin.lee@jpmchase.com}
\authornote{This research was conducted prior to joining JPMorganChase.}


\begin{abstract} 
Desktop environments can integrate augmented reality (AR) head-worn devices to support 3D representations, visualizations, and interactions in a novel yet familiar setting. As users navigate across the dual realities---desktop and AR---a way to move 3D objects between them is needed. 
We devise three baseline transition techniques based on common approaches in the literature and evaluate their usability and practicality in an initial user study ($N$=18). 
After refining both our transition techniques and the surrounding technical setup, we validate the applicability of the overall concept for real-world activities in an expert user study ($N$=6). 
In it, computational chemists followed their usual desktop workflows to build, manipulate, and analyze 3D molecular structures, but now aided with the addition of AR and our transition techniques. 
Based on our findings from both user studies, we provide lessons learned and takeaways for the design of 3D object transition techniques in desktop + AR environments.
\end{abstract}



\begin{teaserfigure}
    \includegraphics[width=\linewidth]{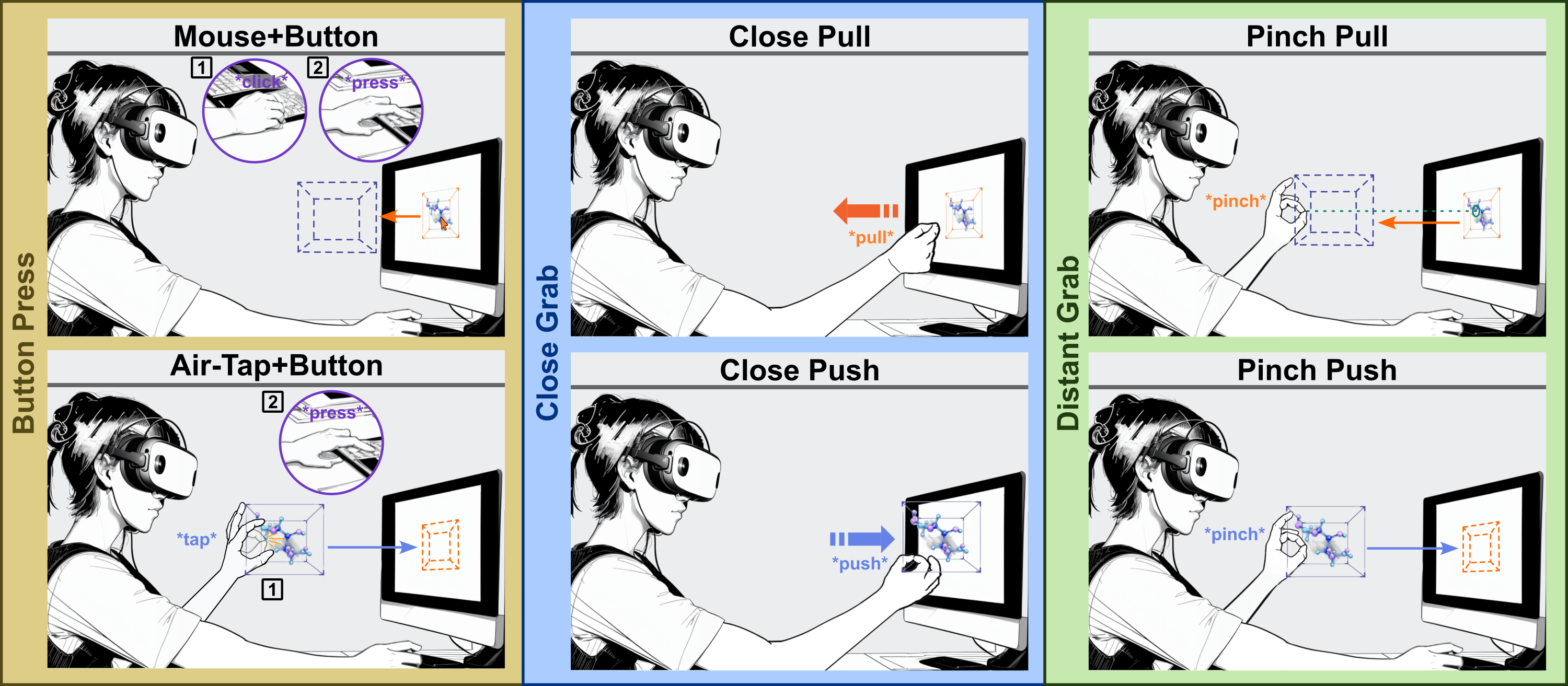}
    \caption{%
    Depiction of our proposed baseline interaction technique pairs ``Button Press'' (\emph{Mouse+Button} and \emph{Air-Tap+Button}), ``Close Grab'' (\emph{Close Pull} and \emph{Close Push}), and ``Distant Grab'' (\emph{Pinch Pull} and \emph{Pinch Push}). Photos were transformed using Stable Diffusion and manually edited and annotated for clarity.
    }
    \Description{
    The figure depicts six different interaction techniques for manipulating objects in a desktop and augmented reality (AR) cross-reality setup. 
    Each technique is demonstrated by a person wearing an AR headset who is interacting with the objects and a monitor. 
    In each technique picture, arrows indicate the trajectory of the virtual object.
    The techniques are categorized in pairs by input type: 
    In the left column is the button press technique pair, which includes the Mouse+Button technique on the top left:
    Clicking a mouse to select an object on the screen is indicated by an inset showing the mouse click.
    Pressing a button to trigger a transition of the virtual object is also indicated by an inset with a finger pressing a button.
    The object comes out of the desktop into the AR environment.
    In the same column, the Air-Tap+Button technique is shown on the bottom left: 
    The person first performs an air-tap gesture to select the object in AR.
    The hand shows a tapping gesture.
    Then, the person presses a physical button to transition the object, indicated by an inset with a finger pressing a button on a keyboard.
    The object goes from the AR into the desktop environment.
    In the middle column, the close grab technique pair is depicted:
    In the picture for the Close Pull technique (Top-middle), a person pulls a virtual object closer while holding a pinch gesture.
    An orange pull arrow indicates the pulling motion toward the person.
    The object is pulled out of the desktop into the AR environment.
    Vice versa, the Close Push (Bottom-middle) technique is used.
    A person pushes a virtual object into the monitor while holding a pinch gesture.
    A blue push arrow shows the pushing motion away from the person.
    The object is pushed from the AR into the desktop environment.
    The right column shows the distant grab techniques:
    The Pinch Pull technique is located at the top right.
    A person targets an object on the monitor and pinches the air with their middle finger.
    The object comes out of the desktop into the AR environment.
    The Pinch Push technique is shown in the bottom right:
    Again, a person targets an object on the monitor and pinches the air with their middle finger.
    The person pinches the air with the middle finger while the middle finger is inside the object to trigger a transition.
    The object goes from the AR into the desktop environment.
    }
  \label{fig:teaser}
\end{teaserfigure}


\maketitle

\section{Introduction}
Digital workflows are shaped by the means which we view and interact with digital content. 2D displays, being the de facto standard, have meant that the majority of digital content and media exists and is portrayed in 2D form: optimized for viewing on a flat rectangular surface. Despite this, there exist many situations in which the constraints of 2D displays become apparent. Most evidently is their intrinsically limited screen real estate and their inability to portray 3D information and structures. 
Augmented and virtual reality (AR/VR) head-mounted displays (HMDs) are thus an increasingly attractive and capable replacement to the 2D displays of old. Their appeal had led to a push towards highly embodied, engaging, and immersive post-WIMP 3D interfaces~\cite{laviola3DUserInterfaces2017}, supplanting the need for 2D displays and their associated interaction metaphors in favor of 3D scenes, content, and visualizations. Yet, we see reimaginings of ``the office of the future''~\cite{grubertOfficeFutureVirtual2018} functionally emulate what we in the 21\textsuperscript{st} century are already familiar with: the same 2D displays represented as virtual panels floating in space around us---further reinforced by companies such as Meta, Apple, and others~(e.g., \cite{bienerQuantifyingEffectsWorking2022}). Unsurprisingly, this spatial computing concept allows users to retain access to the mature ecosystems and input devices of desktop computing within a virtually reconfigurable 3D workspace. While in some ways the ``best of both worlds'', the approach has arguably yet to take full advantage of immersive HMDs capabilities.

Prior research has recognized this by exploring how to more tightly integrate 2D and 3D workflows that go beyond just having 2D panels in 3D. Generally under the moniker of ``cross-reality''~\cite{simeoneInternationalWorkshopCrossReality2020}, these systems allow the use of two or more points of the reality-virtuality continuum (RVC)~\cite{milgram_TaxonomyMixedReality_1994}---in some cases at the behest of the user. For example, the user might go from working on a desktop to putting on a VR HMD to get an equivalent immersive view of their workspace~\cite{schroder_collaborating_2023,hubenschmidReLiveBridgingInSitu2022, aichemHybridUserInterface2022}. This form of \textit{transitional interface}~\cite{billinghurstMagicBookMovingSeamlessly2001}, in effect, incorporates two or more manifestations of the same underlying system (e.g., desktop and VR) that can be switched in an ideally seamless manner, allowing users to choose the interface which best suits their needs. This transition across realities and displays, however, tends to be cumbersome~\cite{rashid_cost_2012} and can be spatially disorientating~\cite{knibbeDreamCollapsingExperience2018,hubenschmidReLiveBridgingInSitu2022}, making their use less compelling in practice.

To ease the transition between realities, it is instead possible to transition digital objects themselves~\cite{wangUserPreferencesInteractive2024, audaScopingSurveyCrossReality2023}, thus keeping the user within the same frame of reference (or \textit{actuality}~\cite{audaScopingSurveyCrossReality2023}). That is, objects are visibly moved from one reality, such as a physical desktop monitor, to another reality, such as in AR, all while the user still perceives themselves to be in AR. This specific hybrid of devices is particularly appealing in instances where an existing desktop workflow is already well established but can be enhanced with AR~\cite{wang_towards_2020}, while still allowing the use of existing desktop input modalities and applications. For example, a user might decide to bring a 3D object or visualization out into AR to get a better sense of its structure, before returning it back to the desktop to continue their work~\cite{lee_design_2022, seraji_hybridaxes_2022, serajiAnalyzingUserBehaviour2024}. In this sense, digital objects are no longer constrained to one device or the other, and can freely move across these \textit{dual realities} at will.

We note, however, that the act of moving an object from the desktop-to-AR (and vice versa) still incurs its own transition cost. 
For instance, consider the visualization and analysis of complex 3D molecular structures on a desktop. 
Not only does the scientist need to specify and trigger the transition, but they also need to maintain their mental model of the 3D structure after the transition is completed---especially should the object's representation change significantly. 
Even seemingly mundane objects that are small and/or visually isotropic (e.g., spheres) still necessitate a non-trivial physical and cognitive effort to both perform the transition and whatever task(s) come afterwards. Whilst previous research has investigated object transitions between desktop and AR environments (e.g.,~\cite{feinerHybridUserInterfaces1991, rekimotoAugmentedSurfacesSpatially1999, wuMegerealityLeveragingPhysical2020, mcdade_realitydrop_2023, coolsDesktopARPrototypingFramework2022, wangUserPreferencesInteractive2024}), none have, to the best our knowledge, properly explored and evaluated the effectiveness of said transition techniques---least of all in real-world use cases.

In this work, we investigate the design of digital object transition techniques across desktop and AR, particularly in single-user contexts where individual objects are transitioned via an explicit action by the user.
We first identify three baseline transition techniques based on those commonly proposed in the literature (\cref{ssc:transfer-techniques}).
We deploy these techniques in an initial user study of 18 participants to understand users' behaviors and preferences between the techniques, and to elicit more general feedback of the desktop + AR setup as a whole (\cref{sc:controlled_study}).
After technique refinement (\cref{sc:technique_refinements}), we validate them in an expert study of six participants in a real-world task: the simulation and analysis of 3D molecular structures (\cref{sc:expert_study}).
We conclude by discussing lessons learned and takeaways for both our transition techniques and for the desktop + AR setting as a whole (\cref{sc:discussion}).

Our contributions are as follows:
\begin{enumerate}
    \item User study ($N=18$) to explore and evaluate three baseline transition techniques
    \item Refined techniques inspired by feedback, observations, and literature
    \item Expert study ($N=6$) to validate the use of the transition techniques in a real-world setting
\end{enumerate}

\section{Related Work}
Our work lies within the field of \textit{cross-reality} (\cref{ssc:cross-reality}), particularly those that combine a desktop with an AR environment (\cref{ssc:screens+ar}). We describe proposed transition techniques from prior work, and their subsequent evaluation where relevant (\cref{ssc:tranisition-digital-objects}).

\subsection{Cross-reality systems} \label{ssc:cross-reality}
The term ``cross-reality'' was coined in 2020 to describe ``the transition between or concurrent usage of multiple systems on the RVC''~\cite{simeoneInternationalWorkshopCrossReality2020}.
Wang and Maurer~\cite{wangDesignSpaceSingleUser2022} proposed a design space specifically for single-user cross-reality applications. Most relevant to us, they identified a common scenario for the \textit{Transition and Concurrent Usage} of cross-reality, that is, the movement of a visualization of one point of the RVC to the other. In this case, it is not the user moving across the RVC---as is the case in \textit{transitional interfaces}~\cite{billinghurstMagicBookMovingSeamlessly2001,carvalhoDesignTransitionalInterfaces2012,georgeSeamlessBidirectionalTransitions2020}---but only the visualization (or object) itself.
Wang and Maurer~\cite{wangDesignSpaceSingleUser2022} also described four forms of interaction in cross-reality systems, two of which---\textit{Moving a visualization across realities} and \textit{Selecting objects across realities}---are necessary to facilitate the aforementioned scenario.
Auda et al.~\cite{audaScopingSurveyCrossReality2023} more recently conducted a survey which identifies \textit{Subsitutional} as one of three main types of cross-reality systems, which follow the principle of having (digital) objects adapt to and be interactable at every available actuality (i.e., an experienceable point of the RVC), such as having a book changing its appearance depending on whether it is seen in reality, AR, or VR~\cite{billinghurstMagicBookMovingSeamlessly2001}. 
Thus, not only can digital objects be moveable across the RVC~\cite{wangDesignSpaceSingleUser2022}, but their appearance and/or interaction affordances should change to match the target actuality~\cite{audaScopingSurveyCrossReality2023} (e.g., from a 2D representation on a monitor to a 3D representation in AR/VR~\cite{lee_design_2022, serajiAnalyzingUserBehaviour2024}).

\subsection{Combining digital screens and AR} \label{ssc:screens+ar}
AR has clear synergy with augmenting digital screens for myriad purposes. The \textit{augmented display}~\cite{reipschlager_designar_2019} is an approach that extends physical screens with 2D and 3D content using AR HMDs. For example, 2D data visualizations can be augmented with 3D annotations and views~\cite{reipschlagerPersonalAugmentedReality2021}, and tablets can have AR visualizations overlaid or aligned next to them~\cite{langner_marvis_2021}. In these works, the AR content is a direct extension of the screen, thus there is little separation between them.
\textit{Complementary interfaces}~\cite{zagermannComplementaryInterfacesVisual2022} instead provide a clearer separation between screen and AR, with user interaction distributed across both devices. For example, a handheld tablet can be used as a control device for AR content~\cite{hubenschmidSTREAMExploringCombination2021,serenoSupportingVolumetricData2019}. In these works, the roles of either device are strictly pre-defined by the system designer.

Our focus on desktop + AR settings is more closely aligned with what Fr\"{o}hler et al.~\cite{frohlerSurveyCrossVirtualityAnalytics2022} consider as ``spatially agnostic cross virtuality,'' which extends a monitor with AR without a strict spatial relationship between the two. The best example (and the largest inspiration behind this work) is that by Wang et al.~\cite{wang_towards_2020} who investigated how 3D visualizations presented in AR can support particle physicists in their existing desktop workflows. Their participants could create and inspect their visualizations both in 2D on the desktop and 3D in AR, which both had equivalent functionality. However, visualizations were authored in either one display or another (with a mechanism to synchronize the two at will), and all interactions were only via mouse input---even in AR. Even still, their findings show great promise in the use of AR to support existing desktop workflows through the use of 3D, and suggest that content in the two environments should be clearly separated without constant synchronization. These insights guide this work into how we might best support the \textit{transition} of digital objects as they move between desktop and AR.

\subsection{Transitioning digital objects between desktop and AR}\label{ssc:tranisition-digital-objects}
When utilizing multiple devices together, a means to transfer data between them becomes vital. Brudy et al.~\cite{brudy_cross-device_2019} proposed a taxonomy for cross-device interactions in ubiquitous computing in which they specified three phases: (1) \textit{configuration} to pair the device(s) that the objects are transferred between; (2) \textit{content engagement} which facilitates the actual transfer, interaction, and exploration of the objects; and (3) \textit{disengagement} to end the pairing. Their literature review shows the influence of device affordances on the interaction design of transfer techniques---particularly the physicality of touchscreens---yet their analysis of AR/VR headsets is minimal, instead relegated to a side category of miscellaneous devices.

Feiner and Shamash~\cite{feinerHybridUserInterfaces1991} is perhaps the earliest example of moving digital content between desktops and AR for meaningful use in both displays, wherein they proposed the use of AR as an extension to the limited display size of the desktop. They described how windows can be dragged out from the screen using the mouse cursor and positioned in 3D space. Similarly, Rekimoto and Saitoh~\cite{rekimotoAugmentedSurfacesSpatially1999} described the \textit{hyperdragging} technique, which allows 2D content such as documents and images to be dragged off the edge of a display using a pointing device, thus transitioning it onto a projected AR surface.
Another early work that describes transition techniques specifically with 3D objects that by Benko et al.~\cite{benko_cross-dimensional_2005}, using a projected tabletop and AR HMD setup. They presented several techniques using hand gestures, including a ``grab and pull'' out from the table and a flat-handed downwards ``push'' back into it. Findings from their user study indicate that such hand gestures were intuitive and easy to perform, likely due to their similarity to real-world interactions. 
More recent research has proposed similar techniques in a proper desktop + AR setting. 
Wu et al.~\cite{wuMegerealityLeveragingPhysical2020} presented techniques for ``pulling'' 3D models from a desktop into AR, though they did not consider the opposite direction. 
McDade et al.~\cite{mcdade_realitydrop_2023} presented three multimodal techniques for the transition of objects both into and out of desktops: Physical Free-hand Drag-and-Drop, Superhuman Hand (for distant transitions), and Superhuman Gaze+Hand. While all three rely on hand gestures, the third incorporates eye gaze to indicate which object on the monitor is to be transitioned. 
Cools et al.~\cite{coolsDesktopARPrototypingFramework2022} proposed a prototyping framework for developing cross-reality desktop + AR systems, envisioning techniques that utilize both mouse and hand input. When transitioning an object from the desktop-to-AR, the mouse serves as the source of the transition in the desktop environment and the hand as the destination in 3D space. The opposite is true when transitioning from AR to the desktop, with the hand as the source and the mouse as the destination. 
Wang et al.~\cite{wangUserPreferencesInteractive2024} instead conducted an elicitation study to determine user preferences for 3D transitions. While they did not directly propose new techniques, they recommend that mid-air hand gestures---mainly drag, tap, and grab---be the primary interaction used to perform them.
All four of these works, however, did not conduct an evaluation of their respective techniques, let alone for real-world tasks.

To the best of our knowledge, the work by Aigner et al.~\cite{aigner_cardiac_2023} is one of few which evaluates the use of object transitions between a screen and AR in an expert pilot study. Cardiologists were asked to plan a surgery in a cross-reality prototype using an isosurface 3D heart model. This model could be transitioned between realities, including a screen and AR, to be viewed in different ways. They received ``overhwelmingly positive'' feedback from their experts, citing the benefits of utilizing multiple systems of the RVC and having seamless transitions between them. While highly relevant, their work used a large display while standing and not seated at a desktop.
Other works such as that by
Seraji et al.~\cite{seraji_hybridaxes_2022, serajiAnalyzingUserBehaviour2024} and Schwajda et al.~\cite{schwajdaTransformingGraphData2023} proposed transition techniques and subsequently evaluated them. However,
these works focus primarily on the visualization domain which requires alternate considerations specific to the field (e.g., the animation of individual components of the visualization~\cite{lee_design_2022}). In this work, we evaluate the usability and practicality of transition techniques in desktop + AR environments, both in generic contexts and in real-world expert contexts.

\section{Context, Prototype, and Transition Techniques} \label{ssc:transfer-techniques}
As any transition technique exists within a broader system and workflow, we first describe the scope to which our techniques are intended to be used in. We then describe the prototype system that our transition techniques were implemented in, followed by the baseline techniques which we later evaluate.

\subsection{Scope}
As cross-reality environments, let alone desktop + AR, can be incredibly broad and wide-ranging, we establish a scope that our chosen techniques are intended for, which we also believe to be the de-facto standard in the literature.

\paragraph{Desktop and AR}
Our focus is primarily on the desktop/monitor (or equivalent) as the 2D space and AR as the 3D space. We choose this as it most closely resembles existing depictions of ``the office of the future''~\cite{grubertOfficeFutureVirtual2018}, and particularly for knowledge workers, working on desktops will likely remain the norm for years to come.

\paragraph{Minimal duplication of objects}
Prior work has indicated that while having the same object be duplicated and synchronized on both desktop and AR can be useful~\cite{coolsDesktopARPrototypingFramework2022, wang_towards_2020}, this duplication should be kept to a minimum~\cite{wang_towards_2020}. Many other works which consider an object as existing only in one environment at a time (e.g.,~\cite{wangUserPreferencesInteractive2024, wuMegerealityLeveragingPhysical2020, benko_cross-dimensional_2005, mcdade_realitydrop_2023, lee_design_2022}).

\paragraph{Close distance from the monitor}
Given the natural affordances of the desktop, most prior work considers transitions when the user is close to the monitor~\cite{wangDesignSpaceSingleUser2022,seraji_hybridaxes_2022,benko_cross-dimensional_2005,coolsDesktopARPrototypingFramework2022,lee_design_2022, wuMegerealityLeveragingPhysical2020,aigner_cardiac_2023}. While McDade et al.~\cite{mcdade_realitydrop_2023} were motivated by the benefits of distant transitions during presentation settings, they did not evaluate any such techniques.

\paragraph{Single-user}
Object transitions are presently considered only in single-user contexts, with multi-user contexts being mentioned either as a motivating use case~\cite{mcdade_realitydrop_2023} or as future work~\cite{schwajdaTransformingGraphData2023,aigner_cardiac_2023}.

\paragraph{Focus on 3D content}
Due to the natural affinity of AR with 3D space, we primarily focus on the manipulation and use of 3D objects across both desktop and AR (e.g., \cite{mcdade_realitydrop_2023, coolsDesktopARPrototypingFramework2022, wangUserPreferencesInteractive2024, wang_towards_2020, wuMegerealityLeveragingPhysical2020, aigner_cardiac_2023}), and not just on 2D windows and information (e.g.,~\cite{feinerHybridUserInterfaces1991,rekimotoAugmentedSurfacesSpatially1999,serajiAnalyzingUserBehaviour2024}).

While our techniques and findings are likely still applicable when used somewhat outside of the scope, such as when using a large display while standing~\cite{schwajdaTransformingGraphData2023, aigner_cardiac_2023}, other techniques would clearly need to be designed if the scenario is vastly different, such as if the monitor is attached to a ceiling and is physically out of reach.

\subsection{Transition Techniques}\label{sc:transiton_techniques}
We now describe our three baseline transition techniques, which intentionally follow designs that, while common in the literature, have yet to be formally evaluated. We focus on keyboard \& mouse~\cite{wang_towards_2020,serajiAnalyzingUserBehaviour2024,coolsDesktopARPrototypingFramework2022} and hand gestures~\cite{mcdade_realitydrop_2023, wuMegerealityLeveragingPhysical2020, wangUserPreferencesInteractive2024, benko_cross-dimensional_2005, coolsDesktopARPrototypingFramework2022, aigner_cardiac_2023} as the two primary input modalities for performing a transition.
Fundamentally speaking, many of these techniques find their roots in existing 3D selection and manipulation techniques (e.g.,~\cite{kang_comparative_2020,jeraldVRBookHumancentered2016,yu_fully-occluded_2020,argelaguet_survey_2013,bergstrom_how_2021,laviola3DUserInterfaces2017, adkins_evaluating_2021}), particularly those of hand and pointer-based interactions.

\begin{figure}[ht]
    \centering
    \includegraphics[width=0.9\linewidth]{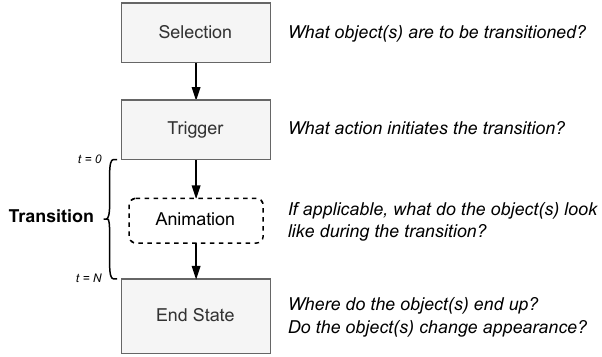}
    \caption{%
    Schematic of a transition process.
    First, the user has to decide which object to transition and make a selection accordingly. 
    Second, a transition has to be triggered.
    Then, a transition is supported by an animation ($t=0$ to $t=N$) until the end state is reached.
    }
    \Description{%
    This figure shows a schematic with boxes, arrows, and text.
    The first is labeled ``Selection'', and next to the box, the question ``What object(s) are to be transitioned?'' is written.
    An arrow goes down from this box to the next box labeled ``Trigger''.
    Again a question is written next to the box ``What action initiates the transition?''
    From this box, there is another arrow pointing to a box that is using a dashed line labeled ``Animations'' and another arrow to the next box.
    The area from the start of the first arrow ($t=0$) to the end of the second arrow ($t=N$) is labeled ``Transition''.
    A question next to this area reads, ``If applicable, what do the object(s) look like during the transition?''
    The last box is labeled ``End State'' with the questions ``Where do the object(s) end up? Do the object(s) change appearance?''
    }
    \label{fig:transition-pipeline}
\end{figure}

To help us better describe our transitions, we introduce a simplified transition pipeline as seen in \cref{fig:transition-pipeline}, which is generally based on the design space of 2D and 3D transformations by Lee et al.~\cite{lee_design_2022} and other related works. It comprises of three logical stages and one optional phase:
\begin{enumerate}
    \item \textbf{Selection.} How does the user denote the object(s) that might later be transitioned? The consideration of interaction methods to indicate a selection before a transition even occurs shows up in many prior works (e.g.,~\cite{mcdade_realitydrop_2023, coolsDesktopARPrototypingFramework2022, wangUserPreferencesInteractive2024})
    \item \textbf{Trigger.} What action does the user perform to initiate the transition? This is the ``essential functionality'' which Wang and Maurer~\cite{wangDesignSpaceSingleUser2022} describe to move an object between realities, and falls under ``content engagement'' by Brudy et al.~\cite{brudy_cross-device_2019}
    \item \textbf{Animation.} Assuming that when the transition is triggered, the object is not instantaneously set to its final transitioned state, how is the object animated throughout this set duration and what does it appear as? While animations are not considered in works involving generic 3D objects (e.g.,~\cite{mcdade_realitydrop_2023, coolsDesktopARPrototypingFramework2022, wangUserPreferencesInteractive2024}), animations have been used when transitioning visualizations between 2D and 3D (e.g.,~\cite{lee_design_2022, schwajdaTransformingGraphData2023})---particularly due to the importance of animations in statistical graphics~\cite{heerAnimatedTransitionsStatistical2007} allowing users to, for example, better track regions of interest~\cite{cordeilAssessingImproving3d2013}
    \item \textbf{End State.} At the end of the transition (which may have included an animation), where do the object(s) end up? Do they change appearance in terms of their position, rotation, scale, or even their material properties and geometry? While a change in position, rotation, and scale (i.e., geometric transform) can be applied to any 3D object, a change in material properties and geometry is highly dependent on the object itself. For example, McDade et al.~\cite{mcdade_realitydrop_2023} showcased how a 3D object can be ``exploded'' into its constituent components~\cite{kalkofenExplosionDiagramsAugmented2009} by transitioning it into AR, yet this is clearly not possible for simpler atomic objects (e.g., a primitive sphere)
\end{enumerate}

We structure our three transition techniques using these stages. However, since each technique is actually a pair of two individual techniques, one in the desktop-to-AR direction and the other in the AR-to-desktop direction, we describe the two separately.
This is because the display and input modality used for the transition naturally influences the design of the technique~\cite{brudy_cross-device_2019}, therefore the techniques in either direction become relatively distinct from each other---especially if they rely on interaction metaphors based on unique device characteristics (e.g., hand tracking). We also align with prior work which also logically separates the two directions~\cite{wangUserPreferencesInteractive2024,lee_design_2022,benko_cross-dimensional_2005}, but still pairing them as complementary techniques~\cite{benko_cross-dimensional_2005}.
Illustrations are shown in \cref{fig:teaser}, and videos are included as supplemental material.
All techniques went through pilot testing with four participants to help adjust any transition parameters. This mainly informed the use of Animations and the chosen End State of the objects after the transition. In particular, all transition techniques include an Animation that interpolates the object from its perceived starting state (i.e., geometric transform) to its defined End State over a duration of 1.5\,seconds with an ease~in~and~out.

\subsubsection{Button Press Technique Pair}
The \textit{Button Press} builds off the notion that whenever the user shifts their attention and workflow between the desktop-to-AR (and vice versa), a switch in input modality is already necessary (e.g., between typing on the keyboard and 3D gesture manipulation). As their hands would likely already be resting on or will soon need to be moved to the keyboard \& mouse, utilizing it as a ``connection point'' between the two displays in the form of a familiar keyboard shortcut is feasible.

\paragraph{Mouse + Button (desktop-to-AR)}
To move an object out of the desktop, a \textbf{Selection} is first made by clicking on the desired object with the mouse~\cite{coolsDesktopARPrototypingFramework2022}, then \textbf{Triggered} by pressing a button on the keyboard (mapped to space bar). After the \textbf{Animation}, the object's \textbf{End State} is set to a fixed position in front of the center of the monitor, which we chose based on pilot study feedback as it was the simplest to understand. The object is rotated to maintain its viewing angle to the user, and is scaled to match the object's ``real'' size to more accurately present it in 3D.

\paragraph{Air-Tap + Button (AR-to-desktop)}
To move an object from AR to the desktop, a Selection is similarly made by air-tapping the desired object with the hand, then Triggered by pressing the same button. The object is then moved to the center of the desktop camera's viewing plane, but keeps its original rotation. The object is automatically scaled to ensure that it remains visible on the screen ($\approx50\%$ screen height), as AR objects can be of sizes much larger than the monitor itself.

\subsubsection{Close Grab Technique Pair}
The \textit{Close Grab} is a standard technique common in the literature (e.g.,~\cite{mcdade_realitydrop_2023, wangUserPreferencesInteractive2024, lee_design_2022, benko_cross-dimensional_2005, serajiAnalyzingUserBehaviour2024, aigner_cardiac_2023}) as it builds directly off direct and embodied manipulation concepts
as well as hand-based 3D interaction techniques (e.g.,~\cite{adkins_evaluating_2021}).
In particular, we consider grab using hand gestures instead of mouse inputs (e.g.,~\cite{coolsDesktopARPrototypingFramework2022}) as gestures can then be used in 3D manipulation tasks.

\paragraph{Close Pull (desktop-to-AR)}
Selection is performed by hovering the hand directly in front of the desired object on the desktop. A pointer appears on the screen to help identify the object that the hand is in front of. The Trigger is when this hand performs a ``pinch'' action with the index finger and thumb. The object is then Animated into an End State with it positioned on the user's hand, billboarded to face the user, and scaled to its real size.

\paragraph{Close Push (AR-to-desktop)}
In the reverse direction, Selection is performed by grabbing the desired 3D object in AR with the same pinch gesture, which is the default in many AR systems. The trigger is when the object is moved by the user such that it collides with the monitor. Based on pilot participant's feedback, we included a radial progress bar as an explicit depth cue for how much further the object needs to be moved towards the monitor, which fills until the two collide. When the transition is initiated, the object is then Animated into an End State with it positioned at the center of the camera's viewing plane, keeping its original rotation, and rescaled to be viewable on the screen.

\subsubsection{Distant Grab Technique Pair}
The \textit{Distant Grab} allows for objects to be transitioned at a distance, again using hand gestures.
This is akin to the many pointer-based selection and manipulation techniques already in the VR literature (e.g.,~\cite{yu_fully-occluded_2020, broussard_evaluation_2024}).
While not necessarily intended to be used on far away screens as McDade et al.~\cite{mcdade_realitydrop_2023} suggested, it prevents users needing to lean forward to be close to the screen as is the case with Close Grab, thus likely reducing physical demand.

\paragraph{Pinch Pull (desktop-to-AR)}
Selection is performed by holding the hand in front of the desired object on the desktop at any distance. A ray perpendicular to the monitor plane helps identify the object that is being hovered over. As pinching with the index finger is often mapped to the grab interaction in AR systems, we instead pinch with the middle finger and thumb to avoid accidental transitions when grabbing other AR objects that are away from the monitor. The transition is Triggered when this pinch is performed. The Animation causes the object to move towards the user's pinching hand, billboarded to face the user, and scaled to its real size.

\paragraph{Pinch Push (AR-to-desktop)}
To transition objects back into the desktop at a distance, Selection is performed by hovering the hand over the desired object, and the Trigger is when the middle finger and thumb pinch gesture is performed on the same hand. The object is then Animated to move towards the center of the camera's viewing plane, keeping its original rotation, and rescaled to be visible on the screen.

\section{Study 1 - Initial Evaluation} \label{sc:controlled_study}
We conducted a user study as an initial exploration and evaluation of the usability and practicality of the proposed transition techniques, and to identify ways they can be refined.
We also investigate if the use of animations have an influence on our observed results, particularly in terms of their position, rotation, and scale.

\subsection{Apparatus} \label{sc:apparatus}
We modified \textit{chARpack}~\cite{rau_charpack_2024, rau_understanding_2024} (The Chemistry AR Package) to be a domain-agnostic basis for our system. 
Using the Unity game engine~\cite{unity}, the prototype is comprised of a desktop client that is wirelessly connected to a standalone AR client (Meta Quest 3 or Microsoft HoloLens 2), using keyboard \& mouse and hand-tracking as input modalities, respectively. 
The baseline functionality of both clients allows for arbitrary digital objects to be freely moved, rotated, and scaled within a 3D coordinate space. The desktop client resembles that of computer-aided design software, allowing its camera to be moved via standard keyboard \& mouse controls.
Our API enables these digital objects and their metadata to be wirelessly transmitted between the two clients, with the receiver reconstructing transitioned objects modified by any additional parameters (e.g., those described in \cref{sc:transiton_techniques}). 
Our transition techniques then leverage this API whenever their respective trigger is met.

Participants sat at a desktop workstation running Windows 10 with a 24\,inch monitor ($1920\times1200$ resolution) and a standard keyboard and mouse. The monitor was positioned 45\,cm from the edge of the desk, allowing for the recommended distance between the monitor and the eye of 52--73\,cm~\cite{rempel_effects_2007}. We opted to use the Meta Quest 3 based on our pilot participants' feedback, mainly due to its higher resolution and field of view.
The livestream of the HMDs perspective was displayed on a separate monitor shown only to the experimenter for observational purposes.
     
\subsection{Tasks}\label{ssc:technique_study_tasks}
\begin{figure}[ht]
    \centering
    \includegraphics[width=\linewidth]{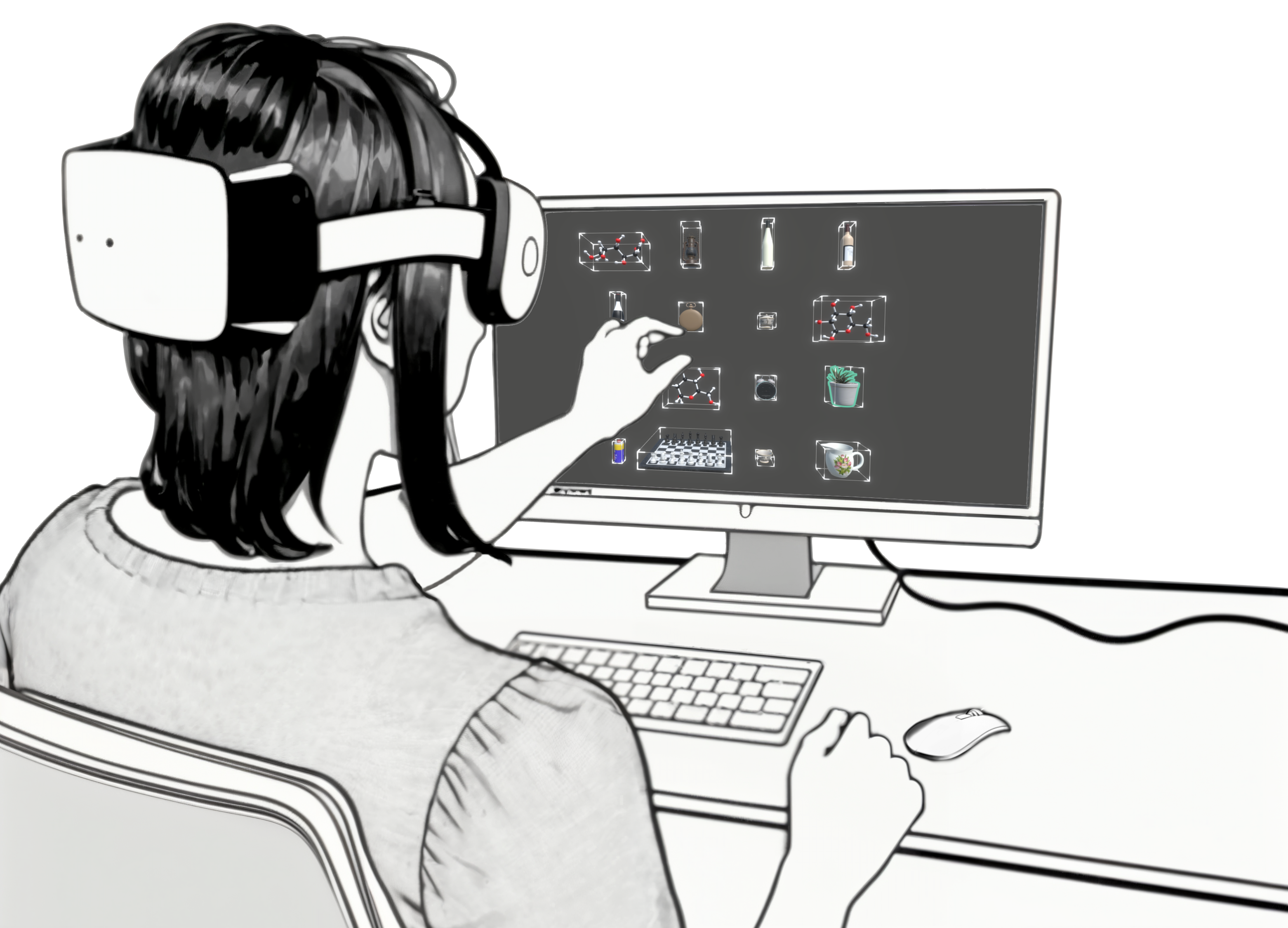}
    \caption{
    The setup for our initial study. 
    The participants sat in front of a monitor wearing an HMD.
    A grid of objects is presented in the desktop environment, and a similar grid is shown in AR next to the participants.
    Image transformed using Stable Diffusion.
    }
    \Description{%
    The picture shows a person from behind.
    The person wears an HMD and sits in front of a desk with a monitor, mouse, and keyboard.
    On the monitor is a grid of virtual objects.
    The person is performing a pinch gesture close to the monitor with the left hand.
    }
    \label{fig:setup}
\end{figure}

All transition techniques were used in a simple docking task that required the use of a transition to complete. 
In each trial, a grid of 16 ($4\times4$) random objects out of a pool of 21 were shown either on the desktop or in AR, which can be seen in \cref{fig:setup}.
For AR, the grid appears on the same side as the participant's handedness within arm's reach of them. 
One random object was highlighted using a blinking turquoise outline, indicating that it is the object that should be transitioned. On the other environment is a duplicate of the same object rendered at 50\% opacity that the transitioned object should be aligned with in its position, rotation, and scale. It was positioned at a random visible location on the desktop or in the field of view in AR, rotated completely randomly, and randomly scaled between 80--120\% of the original object's size.
The task was determined to be complete when the participant pressed the return key on the keyboard, marking the end of the trial.

\subsection{Study Design and Procedure}
We used a within-subjects study design to test our three transition technique pairs as described in \cref{sc:transiton_techniques}.

Participants were first given a brief overview of the study and were asked to sign a consent form. They were then instructed to wear the AR HMD and to sit at the desktop. They were then taught how to move, rotate, and scale objects on the desktop and in AR. The experimenter then walked through each of the six transition techniques, and participants were allowed to spend as much time as they wanted to familiarize themselves with them. This took no longer than 15 minutes.

Participants were then given each of the three transition technique pairs in a counter-balanced order. For each pair, they performed the task described in \cref{ssc:technique_study_tasks} four times: once for each combination of animation (i.e., with and without) and transition direction (i.e., desktop-to-AR or AR-to-desktop). For each trial, we collected task completion time, measured as the time between starting the task and when they pressed the return key, and the number of transitions performed, including those unsucessful (i.e., trigger action was performed but no object was transitioned) and redundant (i.e., repeat transitions of the target object).
After each technique, participants were asked to fill out a questionnaire that consisted of the NASA-TLX~\cite{hart_development_1988}, the System Usability Scale (SUS)~\cite{brooke_sus_1995}, two Likert scale questions regarding the effectiveness of the technique with and without the animation, and free-text feedback regarding the use of animation and the overall technique as a whole. They then repeated this for the remaining~technique~pairs.

At the end of the study, participants were given an additional questionnaire that asked them to rank each technique pair, with free-text feedback regarding subjective preferences and thoughts on the hybrid setup as a whole. Throughout the entire experiment, the experimenter took notes via the livestream of the participant's perspective. This perspective was also recorded alongside with audio for later analysis.

A total of 18 participants (5 female, 13 male) between the ages 18--55 (mode=25--34) were recruited from our university using email distribution lists and word-of-mouth. On a 5-point Likert scale (5 is highest), participants rated their experience with VR and/or AR an average of 3.44 ($SD=1.21$), and their experience with cross-reality interfaces an average of 2.31 ($SD=1.14$).
All participants had normal or corrected-to-normal vision.
All participants except for university employees (due to university policy) were compensated with €12.50 for their time. Each study took 45--60 minutes.

\subsection{Results, Observations, and Feedback}\label{ssc:study_results}

For quantitative measures, we rely on visual analysis using forest plots with the mean and 95\% confidence intervals~\cite{cumming_new_2014, apapublicationmanual}. 
For written feedback and observations, the first two authors conducted a thematic analysis~\cite{braun_using_2006} together on all data, identifying common topics and themes made by participants.

\begin{figure}[ht]
    \centering
    \includegraphics[width=\linewidth]{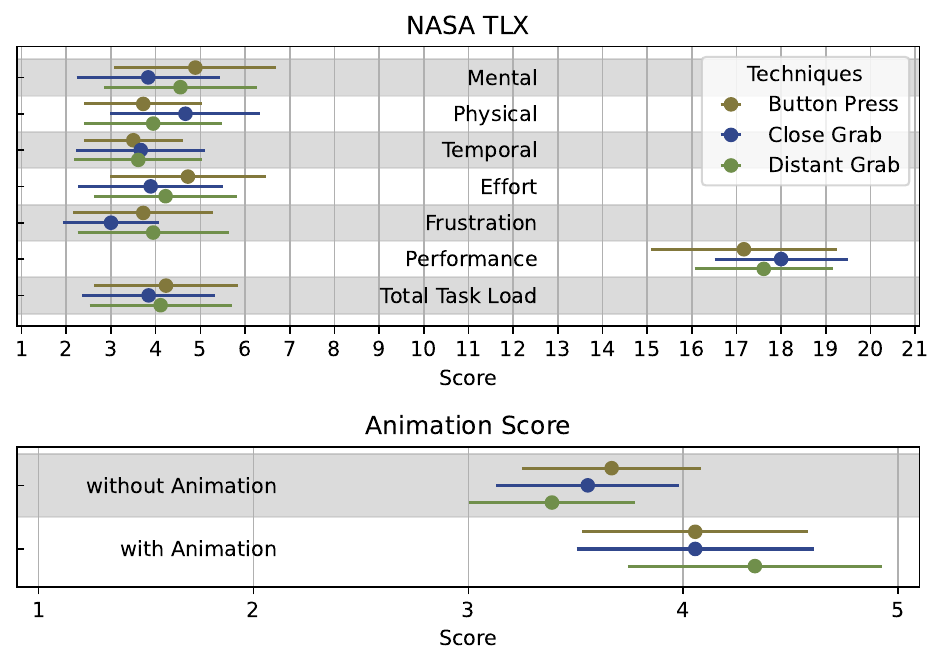}
    \caption{%
    Top: 95\% confidence intervals of the six individual NASA-TLX subscales and the total Raw-TLX score for the transition techniques.
    Bottom: 95\% confidence intervals of Likert scale ratings for the effectiveness of the transition techniques with and without animation.
    }
    \Description{%
    The figure contains two plots: one NASA TLX forest plot on top and an Animation Score Likert-scale forest plot on the bottom.
    Both plots compare the same interaction technique pairs (Button Press, Close Grab, and Distant Grab) by showing mean values and confidence intervals.
    The top forest plot shows the different components of the NASA Task Load Index: Mental, Physical, Temporal, Effort, Frustration, and Performance. 
    An overall Total Task Load is also presented.
    For each workload, the three mean values and confidence intervals of the three technique pairs.
    The scores for each workload component are plotted on the x-axis, ranging from 1 to 21.
    The mean of all techniques and scores, except performance, are located between 3 and 5.
    Notably, the close grab shows the highest physical demand of about 4.7, and the button press was rated with the most mental demand and effort.
    The Performance score shows the button press lowest, then distant grab, and close grab highest, while the values are between 17 and 18.
    All techniques score a total task load closely between 17 and 19.
    On the bottom, the Animation scores are plotted in the same fashion and show the two conditions without Animation and with Animation.
    The scores are plotted on a scale from 1 to 5 on the x-axis.
    Without Animation: The distant grab scored lowest at around 3.4, followed by the close grab, and finally, the button press at about 3.6.
    With Animation: The button press and close grab are almost identical, with a score slightly above 4, and the distant grab has a higher score of about 4.3.
    }
    \label{fig:forest}
\end{figure}

\paragraph{NASA TLX}
The Raw-TLX scores are shown \cref{fig:forest} (top). Overall, there are small differences in task load between the three transition technique pairs. Close Grab performed slightly better on almost all subscales, with the exception of physical demand, reflected by P18 who commented it ``\textit{requires more movement and is thus slower.}''

\begin{table}[ht]
    \centering
    \caption{%
    Aggregated SUS scores, rank scores, and tallies of coded feedback for the transition techniques: Button Press (BP), Close Grab (CG), and Distant Grab (DG). The rank scores are calculated as the sum of all ranks across all participants where best=3, middle=2, and worst=1.
    }
    \label{tab:feedback_scores}
    \begin{tabular}{lccc}
    \toprule
         & \multicolumn{3}{c}{Technique Pairs} \\
         \cmidrule(l){2-4}
         \textbf{Scores} & BP & CG & DG \\
         \midrule
          SUS & 82.08 & 90.56 & 86.39  \\
         Rank & 28 & 44 & 36 \\
         \midrule
         \textbf{Feedback} & & & \\
         Technique positive & 5 & 17 & 12 \\
         Technique negative & 11 & 5 & 13 \\ 
         Technique neutral/suggestion & 1 & 5 & 4 \\
         Did not like/notice animations & 9 & 11 & 3 \\
         Animation is useful/fun & 3 & 5 & 9 \\
    \bottomrule
    \end{tabular}
\end{table}

\paragraph{Aggregated Subjective Feedback}
\cref{tab:feedback_scores} shows the average SUS scores and rank scores for all participants, wherein the rank is calculated as the sum of points across all participants where best=3, middle=2, worst=1. Overall, the close grab was the most subjectively preferred by participants, reflected in both the SUS and ranking. According to Bangor et al.~\cite{bangorDeterminingWhatIndividual2009}, the SUS describes the Button Press as ``Good'' and the Close and Distant Grabs as ``Excellent''. \cref{tab:feedback_scores} also shows the aggregated counts of coded open-ended feedback from our analysis, which indicates that Close Grab received the most amount of positive comments, whilst distant grab and button press were viewed less positively.

\paragraph{Animation}
\cref{fig:forest} (bottom) shows the Likert-scale ratings of the three transition technique pairs both with and without animations. We can observe that all technique pairs see a perceived improvement in subjective rating when animation is included, with Distant Grab benefitting the most. This improvement was also noticeable in the coded feedback, where many participants found animation specifically for the Distant Grab to be useful. This may be because the Distant Grab involves a position change between the start and end points greater than $\approx40$\,cm as compared to the other two technique pairs, to which P15 commented ``\textit{The animation is more helpful because you can see where the object is moving. Without the animation, the object suddenly disappears, which is a bit confusing or surprising.}'' However, eight participants stated that the change in rotation and scale was not useful for the Button Press and Close Grab, though none said the same for the Distant Grab. For example, P12 felt that it was ``\textit{kind of pointless to increase the size, you are going to adjust it yourself anyway,}'' and is reflected in the feedback tallies in \cref{tab:feedback_scores}.

\begin{figure}[ht]
    \centering
    \includegraphics[width=\linewidth]{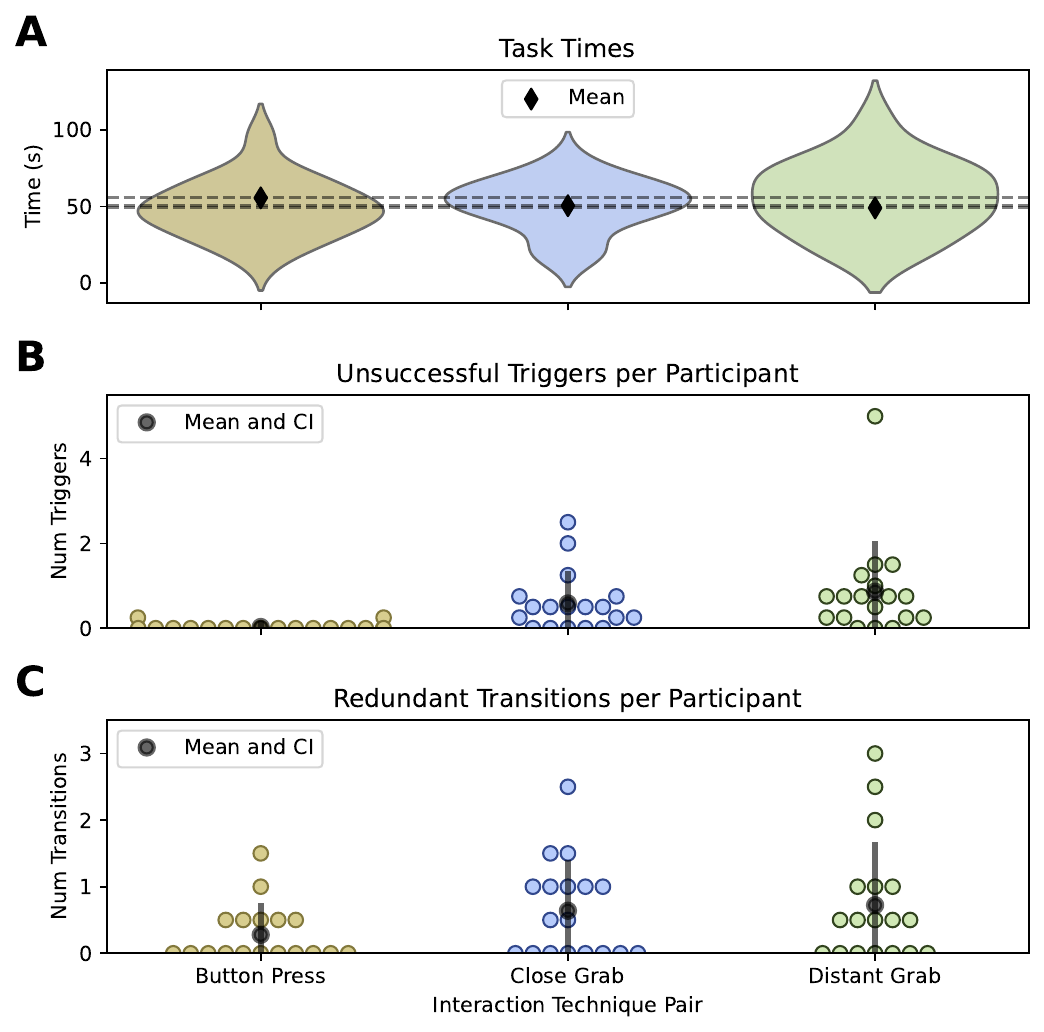}
    \caption{%
    A: Task completion times.
    B: The average number of incorrect triggers performed by each participant per trial and their 95\% confidence intervals.
    C: The average number of redundant triggers performed by each participant per trial and their 95\% confidence intervals.
    }
    \Description{%
    This figure contains three plots: A violin plot for task times per participant (top), a strip plot for unsuccessful triggers per participant (middle), and a strip plot for redundant triggers per participant (bottom).
    All plots compare three different interaction technique pairs: Button Press, Close Grab, and Distant Grab.
    The violin plot shows the distribution of task completion times per participant (in seconds) for each technique pair.
    Button Press (Yellow): Time distribution is wider in the middle, narrowing at the top and bottom, indicating some variation.
    Close Grab (Blue): This plot shows a more compressed distribution, with less variation in task times.
    Distant Grab (Green): Similar to Button Press, but with a more even distribution across times.
    Black diamonds represent the mean task completion times for each technique pair. 
    The means for all three techniques seem to be around 50 seconds.
    A dashed line runs horizontally through the mean task completion times to visually connect the mean values across the techniques.
    The strip plot in the middle shows the number of unsuccessful triggers per participant for each technique pair.
    Each circle represents one participant’s number of unsuccessful triggers, distributed along the y-axis (ranging from 0 to 3 triggers).
    Button Press (Yellow): Only two participants had non-zero unsuccessful triggers.
    Close Grab (Blue): Three participants have an average unsuccessful between 1 and 3; the rest are clustered at the bottom with values between 0 and 1.
    Distant Grab (Green): Most participants are clustered at the bottom with values between 0 and 2; one participant stands out with about 4.5 unsuccessful triggers on average.
    A Black circle with an error bar represents the mean number of unsuccessful triggers and the confidence interval for each technique pair.
    Another strip plot on the bottom shows the number of redundant transitions per participant for each technique pair.
    Each circle represents one participant’s number of redundant transitions, distributed along the y-axis (ranging from 0 to 4.5 transitions).
    Button Press (Yellow): Most participants have under 1 redundant trigger on average, and two participants have values between 1 and 2.
    Close Grab (Blue): A wider spread, with one participant having up to 2.5 redundant triggers on average, but the majority are between 0 and 2.
    Distant Grab (Green): Three participants have between 2 and 3 unsuccessful triggers on average; the rest are clustered very low between 0 and 1.5.
    A Black circle with an error bar represents the mean number of unsuccessful triggers and the confidence interval for each technique pair.
    }
    \label{fig:times_triggers}
\end{figure}

\paragraph{Completion Time and Errors}
Task completion times are shown in \cref{fig:times_triggers} (A). We see all transition technique pairs performing roughly similarly, with some participants taking longer with Distant Grab than the other two techniques. This is evidenced in the number of unsuccessful triggers for Distant Grab made by some participants in \cref{fig:times_triggers} (B), which is likely because the Distant Grab's ``\textit{[middle finger] might lead to an accidental press while doing a pinch.}'' [P16]. Other than this issue, no participants made direct comments on the difficulty in performing the techniques. Interestingly, despite Close Grab having some participants making a comparatively high number of unsucessful trigers, its completion time was still respectable.

\paragraph{General Observations and Behaviors}
We anticipated the task to only require a single transition, with the docking alignment facilitated using the target environment's input modalities. Yet, as can be seen in \cref{fig:times_triggers} (C), we observed many participants transitioning the correct object more than was necessary.
This was mainly to use hand gestures in AR for rotating and scaling the object instead of using the desktop controls, with participants repeating the transition several times until the object aligned with the target object.
Moreover, we observed 11 participants trying to move or rotate objects that had just been transitioned into the desktop using hand gestures. This may be because participants were treating the system not as two distinct environments with a discrete transition between them, but as a singular fluid environment that objects can move around in.
This may also be due to how cumbersome many participants felt it was to have to switch between the keyboard \& mouse and hand gestures, with 11 participants stating it was ``too much effort'' [P1] and that switching was ``somewhat unnecessary'' [P10].
We also observed participants' posture, with eight seated close to the edge of the table and 10 leaning back in a relaxed position, though we did not find a correlation between their posture and their preferred technique.

\paragraph{Outlook and Alternate Techniques}
Participants were generally optimistic towards using a setup similar to our study in the future. Many were able to identify use cases, though almost all involved some form of 3D graphics, such as in scientific visualization [P4], 3D printing [P17], or even for looking at cosmetic items in 3D games like Counter-Strike [P2].
Four participants acknowledged the hardware's limitations regarding the video see-through resolution, but also expressed interest in using an optical see-through display instead to be able to more clearly read the monitor.
Many participants gave direct suggestions for improvement. Four suggested the ability to manipulate objects on the desktop using hand gestures. Three suggested the ability to throw objects into the desktop (akin to that suggested by Wang et al.~\cite{wangUserPreferencesInteractive2024}). The remaining suggested other techniques: closing the eyes as a trigger for the transition [P8], using eye gaze as a selection modality [P18], and using sound as a trigger~[P1].

\subsection{Summary and Takeaways}
Overall, we found that the Close Grab technique pair was the most subjectively preferred despite its higher physical demand and proneness to errors. The use of hand gestures was overwhelmingly preferred, as many participants found the need to swap to the keyboard \& mouse to be unnecessary, even if it had resulted in the lowest errors. Instead, participants expressed their desire to manipulate 3D desktop objects with gestures.
On the other hand, an animated change in position was only deemed necessary when the distance between the start and end of the transition was large---in our study, greater than $\approx40$\,cm. Any changes in rotation and scale generally went unnoticed by participants. Scale, in particular, was seen as unnecessary to automatically adjust. This may, however, be due to the study design, as participants were required to rotate and scale the objects in arbitrary ways; thus, the state of the object after the transition was, in a sense, also arbitrary.

\subsection{Transition Technique Refinements} \label{sc:technique_refinements}
\begin{figure*}
    \centering
    \includegraphics[width=1.0\linewidth]{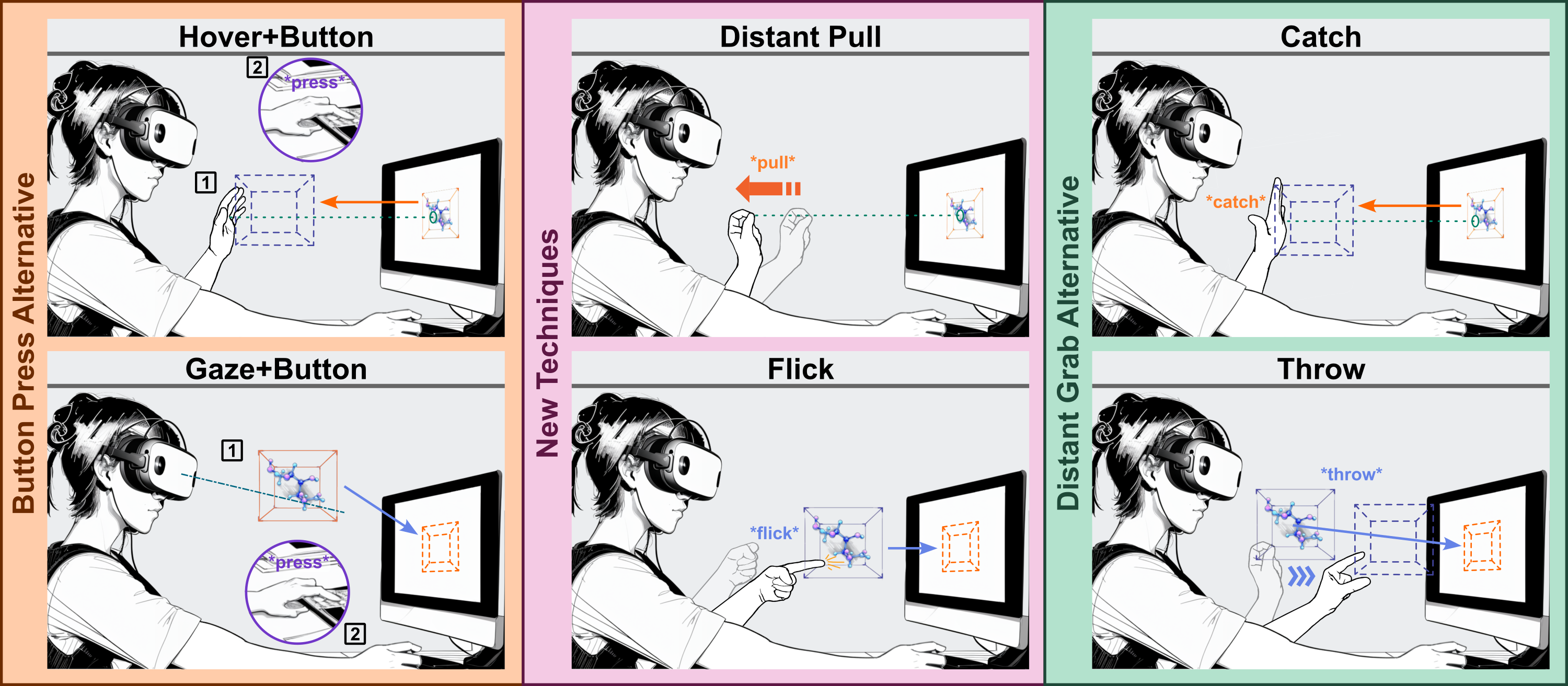}
    \caption{%
    Our refined techniques with the categories ``Button Press Alternative'' (\emph{Hover+Button} and \emph{Gaze+Button}), ``New Techniques'' (\emph{Distant Pull} and \emph{Flick}), and ``Distant Grab Alternative'' (\emph{Catch} and \emph{Throw}). 
    Image transformed using Stable~Diffusion.
    }
    \Description{
    The figure depicts six different interaction techniques alternatives for manipulating objects in a desktop and augmented reality (AR) cross-reality setup. 
    Each technique is demonstrated by a person wearing an AR headset who is interacting with the objects and a monitor. 
    In each technique picture, arrows indicate the trajectory of the virtual object.
    The techniques are categorized in pairs by input type: 
    In the left column are the button press alternative techniques, which includes the Hover+Button technique on the top left:
    Hovering a hand in front of the monitor can be used to select a virtual object.
    Pressing a button to trigger a transition is indicated by an inset with a finger pressing a button.
    The object comes out of the desktop into the AR environment, specifically to the hand of the person.
    In the same column, the Gaze+Button technique is shown on the bottom left: 
    The person is shown with a gaze ray, which is used for the selection of a virtual object.
    Then, the person presses a physical button to transition the object, indicated by an inset with a finger pressing a button on a keyboard.
    The object goes from the AR into the desktop environment.
    In the middle column, new techniques are depicted:
    In the picture for the Distant Pull technique (Top-middle), a person pulls a virtual object from a distance while holding a pinch gesture.
    An orange pull arrow indicates the pulling motion toward the person.
    The object is pulled out of the desktop into the AR environment.
    The Flick technique is shown in the bottom middle.
    A person flicks the index finger on a virtual object, which then moves the object to the monitor and into the desktop environment.
    The right column shows the distant grab alternatives:
    The Catch technique is located at the top right.
    A person targets an object on the monitor and opens their hand to trigger the transition.
    The object comes out of the desktop and to the hand of the person.
    The Throw technique is shown in the bottom right:
    A person throws an object on the monitor to trigger a transition into the desktop environment.
    }
    \label{fig:technique_refinements}
\end{figure*}

We now iterate and refine our transition techniques. We present alternatives to the two worst-performing technique pairs---Button Press and Distant Grab---and introduce new techniques based on participant suggestions. Illustrations of each of these are shown in \cref{fig:technique_refinements}. We opted not to adjust the Close Grab due to it receiving the clearest positive feedback. We also no longer put an emphasis on the change in object rotation and scale, again due to participants not noticing it during the study.

\subsubsection{Button Press Alternative Pair}
As participants' dislike of the Button Press came from needing to switch input modalities, we simplify Selection to minimize the need to click with the mouse or air-tap with the hand prior to the Trigger. We still, however, retain the keyboard button press due to its resistance to accidental triggers and as a representative of hardware input (\cref{fig:times_triggers} B).

\paragraph{Hover + Button (desktop-to-AR)}
Selection is based on a hand hover that shoots a perpendicular ray onto the monitor, selecting the hovered object. When the transition is Triggered with the keyboard button press, the object is Animated towards the hovered hand. This removes the need to use the mouse during this transition.

\paragraph{Gaze + Button (AR-to-desktop)}
Selection is based on the head gaze of the user (not eye gaze due to the Meta Quest 3 not supporting this), selecting the object that is being directly looked at. The Trigger is, again, the button press on the keyboard, with the object Animating towards the center of the screen. This eliminates the need to perform any hand gestures during the transition, instead relying on gaze which was suggested by P18.

\subsubsection{Distant Grab Alternative Pair}
As the main complaint of Distant Grab was the middle finger and thumb pinch gesture, we sought to utilize a different metaphor which was associated with moving objects across large distances.

\paragraph{Catch (desktop-to-AR)}
Similar to related work~\cite{mcdade_realitydrop_2023}, we opt for a ``catch'' metaphor wherein the user holds their palm up in front of the monitor at any distance. Like above, this shoots a perpendicular ray onto the monitor, serving as the Selection method. This also serves as the Trigger for the transition, Animating the object to ``fly'' towards the user's catching hand.

\paragraph{Throw (AR-to-desktop)}
The opposite of catch, we implement the ``throw'' gesture which was requested by our participants and that also appears in related works~\cite{wangUserPreferencesInteractive2024}. The Selection and Trigger are again intertwined, as the transition occurs when the object is in a ``thrown'' state and collides with the monitor, positioning it in the center of the screen. To facilitate this, a grabbed object is considered thrown when it is released at a high velocity, which then imparts the momentum of the hand onto the object.

\subsubsection{On-screen Manipulation and New Techniques}
As many participants either tried or directly requested the ability to manipulate desktop objects using hand gestures, we decided to integrate this functionality into our system. Through this, users can use the conventional pinch gesture to grab and rotate 3D objects on the screen at any distance, aided by the perpendicular ray which acts as a pointer.
This function directly supports the ``Manipulating object across realities" interaction described by Wang and Maurer~\cite{wangDesignSpaceSingleUser2022}.
To complement this, we also devised another transition technique pair which we believed to best complement this feature.

\paragraph{Distant Pull (desktop-to-AR)}
The Selected object is the desktop object currently being grabbed using hand gestures. When the user moves their hand away enough from the screen such that the object collides with the desktop 3D scene camera's near plane, the transition is Triggered, and the object Animated towards said hand.

\paragraph{Flick (AR-to-desktop)}
Following the same concept where a currently grabbed object can be easily transitioned between environments, the user can Select an object in AR simply by grabbing it. The Trigger is therefore doing a ``flick'' gesture~\cite{varshney_flick_2021} with the index finger and thumb, Animating the object to the center of the screen. We note that the flick gesture requires the hand to be in a pinching posture, thus making this a natural two-stage process.

\section{Study 2 - Expert Evaluation}\label{sc:expert_study}
Using our baseline and refined techniques, we then conducted a qualitative expert study with the goal of testing our system in a more realistic scenario.
In particular, we wanted to determine whether our transition techniques could properly support actual workflows in this cross-reality setup as a whole, rather than evaluating the techniques in another controlled setting.
For this, we chose computational chemistry as our domain. 
The typical workflows in computational chemistry interweave tasks of different abstraction, some of them well-suited for the usual desktop setup (like setting up and running numerical simulations), while other tasks will involve visualizing, inspecting, analyzing and manipulating complex 3D data, which in this domain are typically spatial representations of molecular structures. 
These latter tasks often profit strongly from true 3D representations, thus, the use of AR as an extension to the desktop, much as was the case with physicists shown by Wang et al.~\cite{wang_towards_2020}, is fitting.
From a practical standpoint, we had existing connections with computational chemists that made recruiting these participants possible.

\subsection{Participants}
We recruited six computational chemists (1 female, 5 male) who regularly work with 3D molecular structures from our university's chemistry department. One participant had $<$1 year of \textit{computational} chemistry experience, one 1--2, two 2--5, and one $>$5 years. Three participants (E1, E3, E4) had experience with a prior version of the chARpack system that did not include our transition techniques. The other three participants had no previous experience with any VR or AR chemistry software.
All participants had normal or corrected-to-normal vision.

\subsection{Apparatus and Tasks}
We used a similar setup as our initial study with the Meta Quest 3.
We also re-enabled the chemistry-specific functionality that was present in the \textit{chARpack} prototype~\cite{rau_charpack_2024}. In particular, individual atoms, interconnected atoms (i.e., bonds), and entire molecules could be created, re-arranged, and re-configured by the user on either the desktop or AR, which then have inter- and intramolecular forces applied to them, resulting in chemically realistic structures. These structures could be moved between either environment using any of the implemented transition techniques---including the original versions presented in \cref{sc:transiton_techniques}. Required functions were also included, such as tools to run simulations and optimizations on molecules via the command line and configuration files (akin to their existing software), tooltips to see molecular properties, and measurement tools to determine distances and angles between atoms. 
We provided a control panel that allows participants to adjust settings for the Animation and End State, such as the duration and the automatic adjustment of position, rotation, and scale.

Based on the observation that participants could not discern between the physical screen and AR content with the Meta Quest 3, we were also interested in whether an optical see-through AR HMD would be preferred by expert users, especially when used for a real-world task. Thus, we allowed our experts to try both the Meta Quest 3 and Microsoft HoloLens 2 throughout the study to elicit further exploratory and subjective feedback. While this may confound their given feedback, we wanted to observe and evaluate our techniques within a practical, realistic, and holistic scenario, reflecting the fact that in reality, the choice of video or optical see-through HMD can and will differ between users.

Before the study took place, we encouraged our participants to bring their own data, i.e. the molecular structure(s) that they were investigating at the time. We then instructed them to simply use the extended \textit{chARpack} prototype and the implemented transition techniques for their own self-defined tasks.

\subsection{Study Design and Procedure}
We ran our expert study in an exploratory fashion where our participants were free to use the system and its transitions to perform their chosen tasks. Most importantly, we did not restrict which transition techniques they had access to throughout the study, and they were free to use all or none of them as they saw fit.

Participants first received an initial overview of the study and signed a consent form. After sitting at the desktop and putting on the AR HMD of their choosing, they were taught how to use all of the aforementioned functionalities described in \cref{sc:apparatus}---including all of the transition techniques in both \cref{sc:transiton_techniques} and \cref{sc:technique_refinements}. Note that the techniques were not introduced as pairs but as distinct techniques in either transition direction. Participants were encouraged to adjust any Animations and End State settings via the control panel, and were also asked if they wanted to try the other AR HMD. This took approximately 25 minutes.

Participants were then free to use the prototype for their own tasks using the data they had brought. Throughout this stage, the experimenter only intervened when participants asked questions and requested help with how to use its chemistry- or transition-related functions. The experimenter also wrote down observations and notes throughout this stage. This took approximately 40 minutes.

Once each participant felt satisfied in completing their chosen task, they took part in a semi-structured interview where the experimenter asked questions about their thoughts on the different transition techniques, how they adjusted the transition variables, how they felt about the system overall, and any areas for improvement. The interviews were transcribed by the experimenter.

As all of our participants were employees of our university, none were allowed to be compensated for their time. The study took between 70--90 minutes for each participant.

\subsection{Results, Observations, and Feedback}
We now describe our findings which were derived from a thematic analysis approach~\cite{braun_using_2006} on our interview transcripts and notes.

\paragraph{Workflows and using transitions}
Each participant described their workflow during the interviews, which we summarize as follows:
\begin{itemize}
    \item E1, E2, E5: Start with an existing molecule and run simulations on them. During this, periodically check the molecule's 3D structure to evaluate its validity
    \item E3: Start with an existing molecule, split it into two, then align the two molecules to later observe reactions. Then, run simulations on these molecules, periodically checking and adjusting their structures depending on simulation outcomes
    \item E4: Build a molecule from scratch and run it through a geometry optimizer. Then, check the distances between atoms and the angles between bonds, making manual adjustments to generate inputs for further optimization runs
    \item E6: Build molecules from smaller molecules, or align two molecules and run them through an optimizer. Then, check the distances between atoms and angles between bonds. Then, simulate reactions and compare resulting structures
\end{itemize}
Overall, all participants performed tasks, which directly utilized and benefitted from the presence of AR, with them performing an average of 19.83 transitions throughout the study ($SD=5.87$).
We observed them using transitions in either direction (i.e., desktop-to-AR and AR-to-desktop) in equal measure. 
Common reasons for performing a transition were: to bring smaller molecules and fragments that were loaded on the desktop into AR to coordinate and combine them into larger molecules (E1, E2, E3, E5, E6); to bring molecules that were built in AR onto the desktop to then run simulations via the command line (E4); and to bring simulation results from the desktop into AR to then analyze and perform measurements on the resulting molecular structures (E1, E2, E3, E4, E5, E6).
In general, participants stated that they would transition molecules into AR for manipulation and inspection, and move them back to the desktop for saving and further processing.
E5, in particular, did not find much value in viewing the 3D structures in AR. They did clarify that this was due to their honed intuition and familiarity with understanding simpler molecules shown in 2D, but acknowledged the benefits of AR to help understand more complex structures.
Even still, E5 appreciated that ``\textit{since the transition techniques are very simple, this reduces the barrier to even look at structures in 3D.}`` In other words, they would still use the transition techniques as long as it was easy to do so.

\paragraph{Learning transition techniques}
We observed that participants tended to begin their task using only a single transition technique for each direction, but would later expand to using other techniques throughout the study on their own volition. 
For example, E2 only used the Pinch Pull and Pinch Push at the beginning, but eventually tried and used all other techniques. E5 was similar, except starting with the Distant Pull and Throw technique.
E6 was the exception, first using the Pinch Pull and Pinch Push techniques but then expanding only to the Catch, Distant Pull, and Throw techniques. 
Note that these two participants, E5 and E6, had instinctively tried and successfully performed the Distant Pull and Throw techniques during the training phase before the experimenter had introduced them. 
In general, all participants were able to quickly grasp and successfully utilize the transition techniques that they had decided to use, with some even using multiple transitions in quick succession simply for the joy of it. However, E2 took longer than the rest to become accustomed to the AR environment, mainly due to their unfamiliarity with hand-tracking and gestures. 
Everyone agreed that with a bit of practice, the transitions would become second nature to them, especially with E6 noting that ``\textit{getting used to the desktop tools also needed a lot of practice in the beginning.}''

\paragraph{Transition technique preference}
We asked each participant which transition techniques they liked the most and the least.
For most their preferred technique:
E4 and E5 chose the Distant Pull and Throw because it was ``\textit{the most intuitive}'' [E4];
E1 and E3 chose the Close Pull and Close Push also because ``it's \textit{the most intuitive for me}'' [E3];
and E2 and E6 chose the Pinch Pull and Pinch Push because it was ``\textit{the most comfortable}'' [E6] and they could ``\textit{use it from any distance and it's the easiest gesture}'' [E2].
E2 had also commented that they thought the Throw technique was ``\textit{amazing}''.
For least preferred, E2, E4, E5, and E6 all chose the variations of the Button Press, particularly the original Mouse + Button and Air-Tap + Button, as ``\textit{switching input devices is cumbersome}'' [E4].
E1 and E3 did not state a most disliked technique.

\paragraph{Transition and animation adjustment}
We had opted to provide a control panel to allow our participants to adjust the Animation and End State of each technique.
While all participants had changed and experimented with different settings, we observed that participants eventually settled on the same default settings that our techniques had. 
No participants had changed the duration of the animation (default of 1.5\,seconds), and all settled on having the center of the screen and the hand as End State positions after each transition. E6, however, turned off the adjustment of rotation and scale. While we had anticipated doing so would make it difficult to track specific regions of the molecules throughout a transition, E6 did not raise any concern.
Two participants also deactivated specific techniques: E3 disabled the Throw after having accidentally done so whilst repositioning a molecule in AR, and E4 disabled the Catch after triggering it several times.
Nevertheless, all participants agreed that this customization is essential. Five participants stated that in the long term, they would find a consistent set of parameters that works across all transition techniques, with E2 instead adjusting these parameters for each individual technique.

\paragraph{On-screen manipulation}
Being one of the most requested features from our first study, we asked our participants about the usefulness of the gesture-based manipulation objects on the desktop.
All participants except E6 agreed that it was useful, stating ``\textit{I would use that often since I can avoid using the mouse}`` [E4] and ``\textit{I do not have trouble navigating the desktop with mouse and keyboard, but this method is very intuitive}`` [E5].
E6 did not see the need for this feature however, as `\textit{`Instead of manipulating objects on screen, I would transition them into 3D space.}''

\paragraph{Video versus optical see-through}
All participants tried both the Microsoft HoloLens 2 and the Meta Quest 3 as representatives of optical and video see-through devices respectively.
E1 and E2 both preferred the optical see-through, as it did not obstruct the desktop as much and was still sufficient for quickly checking 3D structures.
The remaining participants preferred the video see-through, as it felt more convenient with a much better field-of-view and higher quality 3D rendering. However, E4 suggested that it needed an ``internal screen rendering'' of the desktop to make it easier to read.
In general, all participants commented on device differences in terms of how they influence the overall usability of the setup. They did not comment on the individual transition techniques themselves, indicating that the device choice did not affect the user experience of the techniques enough for our participants to notice or comment.

\paragraph{Envisioned future use of transitions and the system}
We asked participants how they might use the transitions outside of our study setting.
E1 envisioned a scenario where they might analyze and discuss the structure of a molecule with a colleague in AR. They also imagined then being able to throw this molecule to said colleague, who can then transition it onto their monitor and run some calculations on it.
E2, E3, and E6 instead suggested that they might probably prefer to work further away from the monitor, especially to view and analyze very detailed and complex molecules ($\geq500$ atoms) that need to be scaled large enough to inspect.
On the other hand, E5 suggested that they ``\textit{would probably use all techniques, but I usually do not get up from the desk.}''
E4 gave a more straightforward answer: that they would ``\textit{try to stick with one technique that I can master.}''
In general, all participants except E5 would like to integrate a system like ours into their workflow and use it on a daily basis. We note that E5 was the participant with the most experience in computational chemistry.

\section{Discussion \& Limitations} \label{sc:discussion}
Our two user studies provide insight into the design of transition techniques in desktop + AR settings and the future of desktop + AR settings as a whole. We frame our discussion around these two levels, followed by a discussion of the limitations of our work.

\subsection{Designing Transition Techniques}
We now discuss the lessons learned and takeaways regarding the design and usability of our tested transition techniques.
\paragraph{Hand gestures are the prevailing input modality for transitions with 3D objects}
Transition techniques that are reliant on hand gestures received the most positive feedback from our two studies, validating the elicitation study by Wang et al.~\cite{wangUserPreferencesInteractive2024}. While this is naturally due to the emphasis on 3D manipulations in both of our user studies, we note that the related expert study by Wang et al.~\cite{wang_towards_2020} revealed a strong desire from physicists for 3D input, despite the study being intentionally constrained to only a mouse and keyboard. Thus, we can reasonably say that hand gestures should be the primary input modality for transitions, especially to facilitate tasks such as analyzing large 3D objects while away from the desk (as suggested by E2, E3, and E6).
The need to switch modalities was also a common critique in both of our studies, echoing that of prior work~\cite{serajiAnalyzingUserBehaviour2024, wangUserPreferencesInteractive2024}. This is evident by the Button-based techniques being the least subjectively preferred by our participants, though this issue may have been mitigated if the transition trigger was mapped to a mouse input such as the scroll wheel.
Nevertheless, it appears our participants thought of hand gestures as the default modality, with any switches to the keyboard and mouse only being acceptable for when the task demanded it, such as to enter text. This is also evident by the majority of our chemists responding favorably to the on-screen manipulation using hand gestures.
We acknowledge, however, that these findings assume a workflow that is mostly reliant on 3D objects. Workflows consisting of primarily 2D content, such as windows, images, videos, and text, may instil a more 2D-centric mental model of the desktop + AR setting, and thus see mouse \& keyboard-based transition techniques such as mouse drags~\cite{feinerHybridUserInterfaces1991,coolsDesktopARPrototypingFramework2022} more preferred by users. Future work should study whether user preferences differ between these content types, especially if users were to switch between predominantly 3D and 2D tasks regularly (e.g., between analysis and report writing).

\paragraph{Supporting multiple transition techniques at varying distances}
While Button-based transition techniques were the least subjectively preferred, despite having the lowest errors. 
In our expert study, participants were split as to which transition technique they liked best. It seems clear that providing access to multiple transition techniques proved useful, with all of our experts being able to learn and utilize them throughout the study. This was aided by many of the techniques being ``intuitive'' to learn and use.
An obvious reason to have multiple transition techniques beyond just subjective preference is for them to be usable from varying distances, as is the case for 3D selection~\cite{jeraldVRBookHumancentered2016,broussard_evaluation_2024,laviola3DUserInterfaces2017}, especially if the user were to be standing~\cite{wangUserPreferencesInteractive2024} or away from the screen~\cite{mcdade_realitydrop_2023}. Beyond this, many existing 3D selection and manipulation techniques can be adapted as transition techniques, thus providing more options for users to leverage desktop and AR spaces. This includes practical techniques such as multi-selection \cite{argelaguet_survey_2013}, context-aware selection~\cite{zhao_metacast_2024}, and out-of-view target selection~\cite{yu_fully-occluded_2020}, as well as more creative techniques such as performing interactions along image planes~\cite{pierce_image_1997}.

\paragraph{Supporting animations in transitions beyond position, rotation, and scale}
To keep our work as generalizable as possible, we investigated animations only in terms of the position, rotation, and scale of 3D objects. These animations were intended to give visual feedback and help users keep track of objects as they moved between environments~\cite{chalbi_understanding_2018}. We observed that an interpolated position change was the only noticeable animation provided the object was moving across a large enough distance ($\gtrapprox$ 40 cm), with rotation and scale animations going mostly unnoticed. We stress, however, that our statistical power is not strong enough to draw confident conclusions. 
Regardless, a possible reason for this is due to change and inattentional blindness~\cite{mcnamara_perception_2011, rensink_see_1997, kreitz_inattentional_2015}, as participants in the first study may have been instead focused on performing the required interaction and task correctly. 
More likely, however, was that the task was too simple to necessitate animations, with participants being able to quickly identify the position, orientation, and scale of the object post-transition to then complete the docking task using 3D manipulations.
Tasks that involve tracking regions of interest on objects could benefit greater from animations~\cite{heerAnimatedTransitionsStatistical2007, tversky_animation_2002, chevalier_staggering_2014}, though other approaches such as annotations may be sufficient for this purpose.
As mentioned in Section~\ref{sc:transiton_techniques}, however, an object may change in its material and geometric properties, and animation can facilitate the tracking of this change. A notable example of this is in the information visualization domain, whereby animations are used to help keep track of changes in visual glyphs between 2D and 3D states~\cite{schwajdaTransformingGraphData2023,lee_design_2022,lee_deimos_2023}. 
Likewise, other domains utilize 2D graphics that differ significantly from their 3D representations due to abstraction, with both being equally important. This includes 2D blueprints and 3D models in architecture, 2D structural formulas and 3D molecules in chemistry, and 2D circuit diagrams and the physical 3D circuit. Future work may consider how a transition between desktop and AR---and thus a change of abstraction---might be supported by animations. Alternatively, it may be that duplicating both 2D and 3D versions is sufficient~\cite{coolsDesktopARPrototypingFramework2022,wangUserPreferencesInteractive2024}, thus acting as a linked 2D + 3D representation~\cite{hong_survey_2024} with a means to visually connect regions of interest together (e.g,~\cite{prouzeau_visual_2019}).

\paragraph{The use of scale and the composition of transitions}
A notewhile benefit of AR is the ability to visualize objects at their ``real'' scale, which is clearly beneficial in contexts such as furniture design and room planning. While our initial study showed that participants did not notice or care for changes in scale between desktop and AR, this was again likely due to the simple task. Our expert study however elicited interest in manipulating scale for tasks involving very large and complex 3D objects to be inspected in AR at a larger environmental scale~\cite{barba_scale_2011}. This opens up possibilities for egocentric AR perspectives to be used in conjunction with the desktop, such as an (animated) transition from a 2D scatterplot matrix on a screen to a fully immersive 3D scatterplot \cite{kraus_impact_2020}.
That all said, it is possible that scale changes (and by extension position and rotation), should not be coupled as part of a transition technique, but instead as a discrete step after a transition is concluded. Rather than associating a specific transition technique to a specific end state, it may be beneficial to have all transitions end at a comfortable figural scale~\cite{barba_scale_2011} in front of the user, and then have the user manually make position, rotation, and scale adjustments afterwards---not too unlike the docking task in our initial study. While this introduces more operations required by the user, it also provides more fine-grained control which may be more desirable in practice.

\subsection{Designing Desktop + AR Setups}
We now discuss the lessons learned and takeaways regarding the design of desktop + AR setups as a whole.

\paragraph{Cross-reality systems for practical use}
Our expert study highlights the potential of using a cross-reality desktop + AR setup that is supported by object transitions, especially in contexts that rely on 3D information and visualization. For instance, participants commented how the transition techniques allow them to quickly view and make sense of complex 3D structures in AR before returning them back to the desktop.
Our initial use of the Meta Quest 3 had clear drawbacks, particularly the poor text readability on the desktop monitor. We therefore also tested Microsoft HoloLens 2 in our expert study to faithfully elicit feedback, especially as other related research involving AR and screens use optical see-through devices (e.g.,~\cite{langner_marvis_2021,hubenschmidSTREAMExploringCombination2021,mcdade_realitydrop_2023, lee_design_2022, chulpongsatornHoloTouchInteractingMixed2023,wang_towards_2020, coolsDesktopARPrototypingFramework2022}). Despite our small sample size of six experts, it is apparent that neither video or optical see-through, at the time of writing, is superior. Future work may instead seek to evaluate the usability of the desktop + AR metaphor when using virtual monitors (e.g., ~\cite{ozvoldik_yasara_2023}), thus rendering the screens at a higher resolution. While recent work has shown that physical workstations perform better than virtual monitors for single-display tasks, the innate advantages of virtual monitors, when implemented correctly, can far outweigh that of physical monitors~\cite{pavanatto_virtual_2023}. Should the windowing metaphor still be the norm in AR spatial computing contexts, then object transitions may be investigated and specially designed for this purpose, especially as virtual windows can be automatically positioned to make transitions easier.
Ultimately, this assumes the user is always wearing an HMD, which may already be a limiting factor as wearing an HMD for extended periods is still fatiguing~\cite{bienerQuantifyingEffectsWorking2022}. It may be that an approach closer to transitional interfaces, where the user is expected to don or doff the HMD whenever switching between desktop and immersive view (e.g.,~\cite{hubenschmidReLiveBridgingInSitu2022}), be more viable in the short term to see real-world use.

\paragraph{The desktop as a ``window'' into 3D space}
As the manipulation of 3D objects in AR was the core motivation of our work, having the desktop client function similar to 3D content creation software like Blender and Unreal Engine was logical. In this case, as objects on the desktop are also defined in a 3D coordinate system, there are, in effect, two 3D spaces present in such desktop + AR settings. Extending the metaphor, this means that the desktop screen acts as the ``window'' to transition objects into the desktop space, and the desktop camera's viewing plane is the equivalent window to transition objects out into the AR space (e.g., a portal). A practical example of this was demonstrated in our Distant Pull technique, where the object needs to be pulled towards the near plane of the camera for the transition to trigger. 
This notion may have also contributed to our participants' attempts to interact with the desktop space using hand gestures, as it broke down the metaphorical divide between both environments with the two separated only by this window.
This window metaphor has been demonstrated by Aigner et al.~\cite{aigner_cardiac_2023}, who used it as a cutting plane for a 3D visualization that is being transitioned between a screen and AR. A deeper investigation may explore the other potential uses this metaphor might have, or whether having a clearer separation of responsibilities between desktop and AR (e.g., two distinct user interfaces) is still beneficial.
Moreover, the desktop having its own separate camera, while necessary for 3D content applications, may result in situations where desktop objects that need to be transitioned are out-of-view. Because AR is assumed to be present, we imagine techniques like VESAD~\cite{normandEnlargingSmartphoneAR2018} can show these out-of-view objects on the desktop, whether it is to provide a wider field-of-view or to show transitionable objects widgets adjacent to the monitor~\cite{reipschlager_designar_2019, wuMegerealityLeveragingPhysical2020}.

\paragraph{Beyond single-user desktop + AR environments}
While our scope for this work is limited to single-user desktop + AR setups, cross-reality object transitions may be needed with any form of 2D display. This includes different display sizes (e.g., wall-sized displays), shapes (e.g., curved monitors), and mobilities (e.g., handheld). While we believe most if not all of our presented techniques are transferrable to any screen configuration, certain setups would clearly make certain techniques more viable than others. For example, techniques that need to be used close to the screen become impractical on very large displays. 
The same also applies to multi-user scenarios. Should collaborators be sharing the same physical display, techniques like the Close Grab may be awkward to use if one has to lean over their collaborator to use it. Alternatively, novel transition techniques could be designed, such as a cooperative gesture that both users need to perform~\cite{liuCoReachCooperativeGestures2017}. Should collaborators each have their own display, such as the scenario envisioned by E1, the techniques become conceptually similar to that of cross-device transfer~\cite{brudy_cross-device_2019}, with AR serving as the intermediary space that metaphorically connects devices together~\cite{butz_enveloping_1999}. As evidenced by both our study and that by Aigner et al.~\cite{aigner_cardiac_2023} with cardiologists, future research into how these transitions could support collaboration is needed.

\subsection{Limitations}
Our first user study has several limitations which may have influenced our findings. 
First, we treated and tested our initial transition techniques in \cref{sc:transiton_techniques} as pairs. While the desktop-to-AR and AR-to-desktop transitions within each pair were designed to be equivalent, our study design meant that participants could not rate each direction's technique individually thus reducing the granularity of their feedback. Our expert study did not test them as pairs.
Second, the use of the Meta Quest 3's video see-through may have influenced participants' behavior, such as them not realizing an AR object has been transitioned onto the desktop due to poor video quality.
Our expert study sought to avoid this by testing and eliciting feedback for both video and optical see-through HMDs, though newer headsets like the Apple Vision Pro may not have this concern.
Third, the docking task was considered complete when the participant pressed the return key, meaning task completion times were not consistently calculated. This, however, allowed us to observe other behaviors such as participants' further use of transitions into AR to perform 3D manipulations to more accurately align the object with the target. 
Fourth, each condition only had a single trial, making our results less statistically powerful. 
Fifth, both of our studies exhibit a significant gender imbalance, which is a recurring issue in the VR research literature that underrepresents female users and participants~\cite{peck_mind_2020}.

In our expert study, chemistry expertise would clearly be required to interpret molecular properties and potential reaction mechanisms, however, we argue that many of the low-level tasks surrounding this---including our transition techniques---do not require this domain knowledge and have equivalents to any other domain. Therefore, providing a solid degree of generalizability. For example, transitioning a molecule is the same as any other 3D object (see \cref{sc:controlled_study}), coordinating and aligning molecules would be similar to arranging 3D objects together (e.g., level design in 3D games), performing distance and angular measurements in molecules can be done to any 3D object or spatial data (e.g., in architecture), and typing and editing configuration files and command line prompts for molecular simulations is much like typing notes, reports, or scripts (e.g., in knowledge work and programming).

\section{Conclusion}
In this work, we investigated the design and usability of transition techniques in a cross-reality desktop + AR environment. Based on prior work and a pilot study, we devised three baseline techniques, comprised of a desktop-to-AR and AR-to-desktop transition.
We conducted an initial user study of 18 participants to evaluate the three techniques in a simple docking task. Based on these results, we refined and added to our set of transition techniques, testing these in an expert study with six computational chemists using realistic analysis workflows.
We then discussed key lessons learned and takeaways regarding the design of object transitions in desktop + AR environments, including the focus on hand gestures as an input modality, and discussed key points desktop + AR setups as a whole.
We hope this work builds towards making AR a viable extension to existing ways of working on desktops, particularly by making the freeform use of these dual realities much more easier and flexible for people and their work.

\begin{acks}
This work was funded by Deutsche Forschungsgemeinschaft (DFG, German Research Foundation) under Germany’s Excellence Strategy - EXC 2075 - 390740016.
We acknowledge the support of the Stuttgart Center for Simulation Science (SimTech). 
We want to thank our colleagues Kuno Kurzhals, who helped use Stable Diffusion for our figures, and Ayla-Irina Flach and Vivien Schraitle for their contributions to our prototype.
\end{acks}

\section*{Supplemental Material Pointers}
We share our code at \url{https://github.com/KoehnLab/chARpack}.

\section*{Images License and Copyright}
We as authors state that all of our figures are and remain under our own personal copyright, with the permission to be used here. We also make them available under the Creative Commons Attribution 4.0 International (\ccby~CC BY 4.0) license and share them at \url{https://doi.org/10.18419/DARUS-4755}.

\bibliographystyle{ACM-Reference-Format}
\bibliography{abbr, references}

\appendix
\section{Disclaimer}
This paper was prepared for informational purposes by the Global Technology Applied Research Center of J.P. Morgan Chase \& Co. This paper is not a product of the Research Department of J.P. Morgan Chase \& Co. or its affiliates. Neither J.P. Morgan Chase \& Co. nor any of its affiliates make any explicit or implied representation or warranty and none of them accept any liability in connection with this paper, including without limitation, the completeness, accuracy, reliability of the information contained herein and the potential legal, compliance, tax or accounting effects thereof. This document is not intended as investment research or investment advice, or as a recommendation, offer or solicitation for the purchase or sale of any security, financial instrument, financial product or service, or to be used in any way for evaluating the merits of participating in any transaction.
\end{document}